\documentclass[12pt]{article}

\usepackage{epsf}
\usepackage{epsfig}
\usepackage{cite}
\usepackage{amsmath}
\usepackage{graphics}

\begin{document}

\setlength{\textheight}{21.5cm}
\setlength{\oddsidemargin}{0.cm}
\setlength{\evensidemargin}{0.cm}
\setlength{\topmargin}{0.cm}
\setlength{\footskip}{1cm}
\setlength{\arraycolsep}{2pt}

\renewcommand{\thefootnote}{\#\arabic{footnote}}
\setcounter{footnote}{0}

\newcommand{\gtrsim}{ \mathop{}_{\textstyle \sim}^{\textstyle >} }
\newcommand{\lesssim}{ \mathop{}_{\textstyle \sim}^{\textstyle <} }
\newcommand{\rem}[1]{{\bf #1}}
\renewcommand{\thefootnote}{\fnsymbol{footnote}}
\setcounter{footnote}{0}
\def\thefootnote{\fnsymbol{footnote}}

\hfill {\tt April 2017}\\
\vskip .5in

\begin{center}

\bigskip
\bigskip

{\Large \bf Second Octant Favored for Non-Maximal $\theta_{23}$}

\vskip .45in

{\bf  Paul H. Frampton\footnote{paul.h.frampton@gmail.com}} 

\vskip .3in

{Dipartimento di Matematica e Fisica ``Ennio De Giorgi", \\
Universit\`a del Salento and INFN Lecce, Via Arnesano 73100 Lecce, Italy}

\bigskip

\end{center}

\vskip .4in 
\begin{abstract}
One of the most robust relationships predicted by binary tetrahedral
($T^{'}$) flavor symmetry relates the reactor neutrino angle $\theta_{13}$
to the atmospheric neutrino angle $\theta_{23}$, independently of $\theta_{12}$.
It has the form $\theta_{13} = \sqrt{2} |\frac{\pi}{4} - \theta_{23}|$.
When this prediction first appeared in 2008, $\theta_{13}$ was consistent
with zero and $\theta_{23}$ with $\pi/4$. Non-zero $\theta_{13}$
was established by Daya Bay in 2012. Non-zero $|\frac{\pi}{4}-\theta_{23}|$
is now favored by the NO$\nu$A experiment and, for $\theta_{23}$,
the aforementioned
$T^{'}$ relation selects the second octant ($\theta_{23} > \frac{\pi}{4}$)
over the first octant ($\theta_{23} < \frac{\pi}{4}$). This analysis initially
assumes CP conservation in the lepton sector, but leptonic CP violation is  
discussed and it is shown that this specific $T^{'}$ relationship is invariant.

\end{abstract}

\renewcommand{\thepage}{\arabic{page}}
\setcounter{page}{1}
\renewcommand{\thefootnote}{\#\arabic{footnote}}

\newpage

\section{Introduction}

\bigskip
Among the 28 free parameters associated with the standard
model of particle theory, the six mixing angles $\Theta_{ij}$
and $\theta_{ij}$ ($1 \leq i < j \leq 3$) associated respectively
with the quarks\cite{Cabibbo,KM} and the neutrinos\cite{P,MNS}
are the nearest to being calculable in a plausible theory. We
have in mind the use of the binary tetrahedral group $T^{'}$ as a 
family symmetry following the discussion in \cite{FK,Carr,FKM,EFM,EF}.

\bigskip

\noindent
In the present article we shall invoke this family
symmetry involving the binary tetrahedral group $T^{'}$
which, unlike smaller discrete groups such as the
unadorned tetrahedral group $T$, also known as $A_4$, has sufficient structure 
to accommodate and relate both quark and lepton
mixing angles.
In a previous study\cite{FKM},
a quite successful exact formula for the Cabibbo angle $\Theta_{12}$
was derived in a $(T^{'} \times Z_2)$ model 
arriving at a Cabibbo
angle value at the lowest order given by
\begin{equation}
\tan 2(\Theta_{12})_{T^{'}}  = \left( \frac{1}{3} (\sqrt{2}) \right).
\end{equation}
\label{CabibboTprime}

\noindent
The three neutrino mixing angles $\theta_{12}, \theta_{23}, \theta_{13}$
have also been studied assiduously in a $T^{'}$ context and
the most robust prediction \cite{EFM} is that

\begin{equation}
\theta_{13} = \sqrt{2} \left| \frac{\pi}{4} - \theta_{23} \right|
\label{root2}
\end{equation}
which interestingly links the non-zero value for $\theta_{13}$
to the departure of the atmospheric neutrino mixing angle
$\theta_{23}$ from maximal mixing with $\theta_{23} = \pi/4$.
This is the most definite such prediction from $T^{'}$,
independent of any further phenomenological input
\footnote{A similar prediction from a different
starting point appeared {\it en passant} in \cite{HS}. Eq.(\ref{root2})
was later derived in \cite{King1}}.

\bigskip

\noindent
Subsequent to the first appearance\cite{EFM} of Eq.(\ref{root2}),
a further analysis in \cite{EF}
found consistency of Eq.(\ref{root2}) with experiment, although at that time
the RHS of Eq.(\ref{root2}) was empirically indistinguishable from zero
so the comparison between theory and experiment was incomplete.
In the present article, we study further the implications of Eq.(\ref{root2}).

\bigskip

\noindent
Recent more accurate data from NOvA \cite{NOvA}, Daya Bay \cite{DayaBay}
and Double Chooz \cite{DoubleChooz} - especially the preference
for nonmaximal $\theta_{23}$ from NOvA data - make possible this more detailed
check.

\bigskip
\bigskip

\newpage

\section{Experimental Data} 

\noindent
The most suggestive data on nonmaximal $\theta_{23}$ comes from the
measurement\cite{NOvA} of $\nu_{\mu}$ disappearance at NOvA as
observed by the near and far detectors. This
experiment finds a new value for the squared mass difference 
$\Delta m_{23}^2$ which is

\begin{equation}
\Delta m_{23}^2 = (2.52^{+0.20}_{-0.18}) \times 10^{-3}  eV^2
\label{m23squared}
\end{equation}

\bigskip

\noindent
The statistically-favored best fit of NOvA for a normal hierarchy (NH) give

\begin{equation}
\sin^2 \theta_{23} = 0.43 ~~ {\rm or} ~~ 0.60 ~~~~~ {\rm (NH)}
\label{NH}
\end{equation}

\noindent
while for an inverted hierarchy (IH) the NOvA best fits are

\begin{equation}
\sin^2 \theta_{23} = 0.44 ~~ {\rm or} ~~ 0.59 ~~~~~ {\rm (IH)}
\label{IH}
\end{equation}

\bigskip

\noindent
The first and second solutions in Eqs.(\ref{NH}) and (\ref{IH}) correspond to the
first octant ($\theta_{23} < \pi/4$) and the second octant 
($\theta_{23} > \pi/4$) respectively. In this note we attempt to
discriminate between these two octants for $\theta_{23}$ by
using the $T^{'}$ relation, Eq.(\ref{root2}) above.

\bigskip

\noindent
To carry this out, we need the best data on $\theta_{13}$ which are
from two experiments \cite{DayaBay,DoubleChooz}. From neutron
capture on Hydrogen $H$ the result found at the Daya Bay reactor in Guangdong
Province, China is
\begin{equation}
\sin^2 2\theta_{13} = 0.071 \pm  0.010 ~~~~~ {\rm (nH)}
\label{DayaBay}
\end{equation}

\bigskip

\noindent
The best value found using the Double Chooz reactor experiment in Chooz, France from neutron
capture on Gadolinium $Gd$ is

\begin{equation}
\sin^2 2\theta_{13} = 0.088 \pm  0.033 ~~~~~ {\rm (nGd)}
\label{DoubleChooz}
\end{equation}

\bigskip

\noindent
When we convert Eqs.(\ref{DayaBay}) and (\ref{DoubleChooz}) to
degrees for comparison with the nonmaximailty of $\theta_{23}$,
we find for the central values in a form convenient to check Eq.(\ref{root2})
\begin{equation}
\left ( \frac{\theta_{13}}{\sqrt{2}} \right)_{nH} = 5.46^{0}
\label{nH}
\end{equation}

\noindent
and

\begin{equation}
\left( \frac{\theta_{13}}{\sqrt{2}} \right)_{nGd} = 6.10^{0}
\label{nGd}
\end{equation}

\noindent
respectively.

\newpage

\noindent
These results are compared to the nonmaximality of $\theta_{23}$ 
cited earlier in Eqs.(\ref{NH}) and (\ref{IH}) within the following table

\bigskip

\begin{table}
\caption{Values of $|\pi/4 - \theta_{23}|$ corresponding to the NOvA data.}
\begin{center}
\begin{tabular}{||c|c|c|c|c||}
\hline
Hierarchy & $\sin^2 \theta_{23}$  & $\sin\theta_{23}$ & $\theta_{23}$ & $|\pi/4 - \theta_{23}|$   \\
  &  &  &degrees&degrees \\
\hline
\hline
NH & 0.43 & 0.66 & 40.97 &  4.03  \\
\hline
NH  & 0.60 & 0.77 & 50.77 & 5.77 \\
\hline
IH & 0.44 & 0.66 & 41.55 & 3.45 \\
\hline
IH & 0.59  & 0.77 & 50.18 & 5.18 \\
\hline
\end{tabular}
\end{center}
\label{longevity}
\end{table}

\bigskip

\noindent
When we compare the values of $\theta_{13}/\sqrt{2}$ given in Eqs.(\ref{nH})
and (\ref{nGd}) we note that only the second octant solution ($\theta_{23} > \pi/4$)
is consistent. The first octant solution ($\theta_{23} < \pi/4$) is not.
\footnote{Note that in \cite{EF} only the first octant for $\theta_{23}$ was
considered because at that time the data were insufficently accurate to
discriminate between octants.}

\newpage

\section{CP Violation}

\noindent
The above analysis assumed that no CP violation is associated with the
neutrino mixing. The experimental situation about this is in a state
of flux\cite{NOvA,PDG2016,NOvA2,T2K}, but it is not premature to discuss how 
or whether non-vanishing $\delta_{CP}$ could affect the $T^{'}$ relation, Eq.(\ref{root2}).
The experimental papers \cite{NOvA,PDG2016,NOvA2,T2K} are consistent at
$3\sigma$ with CP conservation, $\delta_{CP} = 0$, but there is a {\it hint} from the data
suggestive of a phase $\delta_{CP} \simeq 3\pi/2$ at a level between
$1\sigma$ and $2\sigma$ so let us re-do the $T^{'}$ analysis 
assuming general $\delta_{CP} \neq 0$.

\bigskip

\noindent
This requires us to re-examine the original derivation of Eq.(\ref{root2})
in the EFM paper \cite{EFM}. We retain the convenient notation 

\begin{equation}
\theta_{ij} \equiv \left( \theta_{ij} \right)_{TBM} + \epsilon_k
\label{epsilon}
\end{equation}

\noindent
where
\begin{equation}
\left( \theta_{12} \right)_{TBM} = \tan^{-1} \left( \frac{1}{\sqrt{2}} \right) ~~~ 
\left( \theta_{23} \right)_{TBM} = \left( \frac{\pi}{4} \right)
~~~ \left( \theta_{13} \right)_{TBM} = 0
\label{TBMvalues}
\end{equation}

\noindent
are the Tri-Bi-Maximal (TBM) values. Our aim is to relate $\epsilon_1$ and $\epsilon_2$
by perturbation around the $T^{'}$ value for the Cabibbo angle $\Theta_{12}$
\begin{equation}
\tan \left[ 2 \left( \Theta_{12} \right)_{T^{'}} \right]  = \left( \frac{\sqrt{2}}{3} \right)
\label{Cabibbo}
\end{equation}

\bigskip

\noindent
The PMNS mixing matrix, including non-zero $\delta_{CP}$, is {\it e.g} \cite{King}
\begin{equation}
\left( \begin{array}{ccc}
- s_{12}s_{23} - c_{12}c_{23}s_{13} \exp(i\delta_{CP}) &  
-s_{12}c_{23}+c_{12}s_{23}s_{13}\exp(i\delta_{CP})    &    
c_{12}c_{13}    \\
c_{12}s_{23} - s_{12}c_{23}s_{13} \exp(i\delta_{CP})  &
c_{12}c_{23}+s_{12}s_{23}s_{13} \exp(i\delta_{CP})     & 
 s_{12}c_{13}   \\
 c_{23}c_{13}  & -s_{23}c_{13}   &  s_{13} \exp(-i \delta_{CP})    
   \end{array}
   \right)
   \label{PMNS}
   \end{equation}
   
\noindent
Substituting Eq.(\ref{TBMvalues}) into Eq.(\ref{PMNS}) gives, for any
value of $\delta_{CP}$,

\begin{equation}
U_{PMNS}(TBM,\delta_{CP}) 
= \left( \begin{array}{ccc}
- \sqrt{\frac{1}{6}} & -\sqrt{\frac{1}{6}} & +\sqrt{\frac{2}{3}} \\
+\sqrt{\frac{1}{3}} & +\sqrt{\frac{1}{3}} &  +\sqrt{\frac{1}{3}} \\
+\sqrt{\frac{1}{2}}  &  -\sqrt{\frac{1}{2}}  &  0  
\end{array}
\right)
\label{TBMmatrix}
\end{equation}

\bigskip

\newpage

\noindent
Likewise, inserting Eq.(\ref{TBMvalues}) into Eq.(\ref{epsilon}) gives, independently
of $\delta_{CP}$ and assuming only that $|\epsilon_i| << 1$,

\begin{equation}
s_{12} =\sqrt{\frac{1}{3}} (1 + \sqrt{2} \epsilon_3) ~~~ c_{12} = \sqrt{\frac{2}{3}}(1-\epsilon_3 / \sqrt{2})
\end{equation}
\begin{equation}
s_{23} = \sqrt{\frac{1}{2}} (1 + \epsilon_1) ~~~ c_{23} = \sqrt{\frac{1}{2}} (1 - \epsilon_1)
\end{equation} 
\begin{equation}
s_{13} = \epsilon_2 ~~~ c_{13} = 1
\end{equation}

\noindent
Using the small $\epsilon_i$ approximation we may write
\begin{equation}  
U  =U_{TBM} +\delta U_1 \epsilon_1 +\delta U_2 \epsilon_2 +
\delta U_3 \epsilon_3  
\end{equation}

\noindent
in which

\begin{eqnarray}
\delta U_1 & = & \left( \begin{array}{ccc}
-\sqrt{\frac{1}{3}}  &  + \sqrt{\frac{1}{3}}  &  0  \\
- \sqrt{\frac{1}{6}}  &  + \sqrt{\frac{1}{6}}  &  0  \\
0 &  0  & 0 
\end{array}
\right)
\label{pert1}
\end{eqnarray}

\begin{eqnarray}
\delta U_2 & = & \left( \begin{array}{ccc}
-\sqrt{\frac{1}{3}}  &  + \sqrt{\frac{1}{3}}  &  0  \\
- \sqrt{\frac{1}{6}}  &  + \sqrt{\frac{1}{6}}  &  0  \\
0  &  0  & 0 
\end{array}
\right)
\label{pert2}
\end{eqnarray}

\begin{eqnarray}
\delta U_3 & = & \left( \begin{array}{ccc}
-\sqrt{\frac{1}{3}}  &  - \sqrt{\frac{1}{3}}  &  -\sqrt{\frac{1}{3}}  \\
- \sqrt{\frac{1}{6}}  &  - \sqrt{\frac{1}{6}}  &  +\sqrt{\frac{2}{3}}   \\
0  &  0 & 0 
\end{array}
\right)
\label{pert3}
\end{eqnarray}

\noindent
For TBM mixing, before CP violation associated with non-vanishing $\theta_{13}$,
one has (with everything still real):

\begin{equation}
\left( M_{\nu} \right)_{TBM} = U_{TBM}^T (M_{\nu})_{diag} U_{TBM}
\label{perturbation}
\end{equation}
\noindent
where
\begin{equation}
(M_{\nu})_{diag} = \left( \begin{array}{ccc} m_1 & 0 & 0 \\
0 & m_2 & 0 \\
0 & 0  & m_3 
\end{array}
\right)
\end{equation}

\newpage

\noindent
This, in turn, yields a formula for $(M_{\nu})_{TBM}$. Defining
$m_{12}=(m_1-m_2)$ the symmetric matrix is
\begin{equation}
\left( M_{\nu} \right) = \left(\frac{1}{6} \right)
\left( \begin{array}{ccc}
m_1+m_2+3m_3 & m_1 + 2m_2 - 3m_3 & -2m_{12}  \\
      &  m_1+2m_2 +3m_3  &   -2m_{12} \\
      &      &   4m_1 + 2m_2  
       \end{array}    \right)
      \end{equation}

\noindent
It follows from Eq.(\ref{perturbation}) that the perturbation is given by
(also using the notations of Eqs.(\ref{pert1}) to (\ref{pert3}))
\begin{eqnarray}
\delta \left(M_{\nu}\right)_{TBM} & = & \left(\begin{array}{ccc}
\delta m_1 & 0 & 0 \\
0 & \delta m_2  &  0 \\
0  & 0 & \delta m_3  
\end{array}
\right)   \\
& = & \delta U (M_{\nu})_{TBM} U_{TBM}^T \nonumber \\
&  & +  U_{TBM} \delta M_{\nu} U_{TBM}^T \nonumber \\
& & + U_{TBM} (M_{\nu})_{TBM} \delta U^T
\label{deltaMnu2}
\end{eqnarray}

\bigskip

\noindent
At this stage, we must ensure that the $T^{'}$ calculation is suitably
generalized to include CP-violating Yukawa couplings,
meaning that the $Y_i$ in the leptonic lagrangian

\begin{eqnarray}
{\cal L}_Y & = & \frac{1}{2} M_1 N_R^{(1)} N_R^{(1)} +M_{23} N_R^{(2)} N_R^{(3)} \nonumber  \\
& &  +Y_1 \left(L_L N_R^{(1)} H_3 \right) \nonumber \\
& &  +Y_2 \left(L_L N_R^{(2)} H_3 \right) \nonumber \\ 
& &   +Y_3 \left(L_L N_R^{(3)} H_3 \right)
\label{lagrangian}
\end{eqnarray}

\noindent
are now complex, unlike in earlier $T^{'}$ discussion. The Majorana masses
of the right-handed neutrinos are real.

\bigskip

\noindent
Just to recall that, under the group $(T^{'} \times Z_2)$ the Higgs doublet
$H_3$ is a $(3, +1)$. We shall not need the charged lepton mass terms
which have been omitted in Eq.(\ref{lagrangian}).
The VEV is $<H_3> = (V_1, V_2, V_3)$, with real $V_i$ because CP
is preserved by the vacuum. 

\bigskip

\noindent
The right-handed neutrinos $N_R$ have a real mass matrix

\begin{equation}
M_N = \left( \begin{array}{ccc}
M_1  &  0  &  0 \\
0 & 0 & M_{23} \\
0 & M_{23} & 0 
\end{array}
\right)
\label{NR}
\end{equation}

\noindent 
while the Dirac mass matrix is complex
      
\begin{equation}
M_D = \left( \begin{array}{ccc}
Y_1 V_1  &  Y_2 V_3 &  Y_3 V_2 \\
Y_1 V_3  &  Y_2 V_2  &  Y_3V_1 \\
Y_1 V_2  & Y_2 V_1  &  Y_3 V_3
\end{array}
\right)
\label{Dirac}
\end{equation}

\noindent
because of the $Y_i$. Next we implement the see-saw mechanism\cite{Minkowski}

\begin{equation}
M_{\nu} = M_D M_N^{-1} M_D^{\dagger}
\end{equation}

\noindent
and define auxiliary variables $x_1 \equiv |Y_1|^2 /M_1$ (real)
and $x_{23} \equiv Y_2^* Y_3 /M_{23}$ (complex) so that $M_{\nu}$ becomes
the necessarily symmetric matrix
\begin{equation}
\left( \begin{array}{ccc}
x_1 V_1^2 + 2x_{23} V_2 V_3  & x_1 V_1 V_3 + x_{23}(V_1^2 + V_1 V_3) &
x_1V_1V_2 + x_{23}(V_3^2 + V_1 V_2)   \\
  &  x_1 V_3^2 + x_{23}V_1V_2 & x_1V_2V_3 + x_{23}(V_1^2 + V_2V_3)  \\
  &  &  x_1V_2^2 + 2x_{23}V_1 V_3
  \end{array}
  \right)
  \label{Mnu}
  \end{equation}
  
  \noindent
  The next step is to compare the matrix (\ref{Mnu}) with the complex texture
  
\begin{equation}
\delta M_{nu} = V_1^{'2} x_1 \left( \begin{array}{ccc}
  2(-2a+b)y & a+(a-4b)y & b+(2a+b)y \\
    &  2(a+by) & (-2a+b)(1+y) \\
      &  & -4b+2ay 
      \end{array}
      \right)
      \label{deltaMnu}
      \end{equation}
      
      \noindent
      which is obtained by using the perturbed VEV 
      
      \begin{equation}
      < H_3 > = V_1^{'} (1, -2+b, 1+a)
      \end{equation}
      with $a, b<<1$ both real. The key observation is that only the (complex)
      parameter $y = x_{23}/x_1$ survives, despite the fact
      that the Yukawa couplings and the right-handed neutrino masses are
      all empirically unknown. It is now straightforward but tedious algebra to 
      eliminate the unknown parameter $y$ from Eq.(\ref{deltaMnu2}) and
      Eq.(\ref{deltaMnu}) to obtain six independent equations, only one of 
      which is relevant to our present discussion. It is      
      \begin{equation}
      \epsilon_2= \sqrt{2} |\epsilon_1|
      \label{result}
      \end{equation}
      
      \noindent
      so that, perhaps surprisingly, for arbitrary $\delta_{CP}$ the $T^{'}$ formula,
      \begin{equation}
      \theta_{13} = \sqrt{2} |\pi/4 - \theta_{23}|
      \end{equation}
      stated near the beginning as Eq.(\ref{root2}), first derived for $\delta_{CP} =0$ in \cite{EFM},
      remains invariant even in the presence CP violation with $\delta_{CP} \neq 0$.
  
\newpage

\section{Discussion}

\noindent
At first non-trivial order the prediction of $T^{'}$ flavor symmetry 
for the PMNS angles $\theta_{ij}$ are

\begin{equation}
\tan^2 \theta_{12} =  \left( \frac{1}{\sqrt{2}} \right)  ; ~~ {\rm and} ~~ \theta_{13} = \sqrt{2}\left| \frac{\pi}{4} - \theta_{23} \right|
\label{Tprimeneutrino}
\end{equation}

\noindent
and the $\theta_{12}$ prediction is consistent within less than $2\sigma$
of the latest data\cite{PDG2016} which averages to $\sin^2\theta_{12} = 0.304 \pm 0.014$.
The second formula in Eq(\ref{Tprimeneutrino}), for $\theta_{13}$, remains
invariant in the presence of CP violation.

\bigskip

\noindent
Our main result here is that combining the latest data on $\theta_{23}$
and $\theta_{13}$ with the $T^{'}$ relation, Eq.(\ref{root2}), favors that
the atmospheric neutrino angle $\theta_{23}$ lies in the second rather
than the first octant, a distinction which cannot be made from the experimental
data alone.

\bigskip

\noindent
It will be very interesting to discover how future more precise measurements
of the neutrino mixing angles remain consistent with these predictions.

\vspace{3.0in}

\begin{center}

\section*{Acknowledgements}

\end{center}

\noindent
This work was initiated at the MIAMI2016 Conference held in Fort Lauderdale
in December 2016 and organized by T.L. Curtright of the University of Miami.

\bigskip
\bigskip
\bigskip
\bigskip

\end{document}